\documentclass[12pt]{article}

\usepackage{vmargin}
\setmarginsrb{3cm}{2cm}{3cm}{2cm}{1cm}{1cm}{1cm}{1cm}
\usepackage{amsmath}
\usepackage{amssymb}
\usepackage[dvips]{graphicx}
\usepackage{rotating}
\usepackage{psfrag}
\usepackage{fancyhdr}
\def\bra#1{\langle #1 |}

\def\ket#1{| #1 \rangle}

\newcommand{\va}{\scriptscriptstyle}
\newcommand{\vani}{\scriptstyle}
\author{Karim Noui\thanks{karim@phys.univ-tours.fr} \\
{LMPT, Parc de Grandmont, 37200 Tours, FR}\\ \\
Alejandro Perez \thanks{perez@cpt.univ-mrs.fr} \\
{CPT, Campus de Luminy, 13288 Marseille, FR}\\ \\
Kevin Vandersloot \thanks{kfvander@gravity.psu.edu}\\
{ICG, University of Portsmouth, Portsmouth PO1 2EG, UK}}
\title{\bf Cosmological Plebanski Theory}
\date{\today}

\begin{document}

\maketitle

\begin{abstract}
We consider the cosmological symmetry reduction of the Plebanski action as a toy-model to explore,
in this simple framework, some 
issues related to loop quantum gravity and spin-foam models.
We make the classical analysis of the model and perform both path integral and canonical quantizations.
As for the full theory, the reduced model admits two disjoint types of classical solutions: topological and 
gravitational ones. The quantization mixes these two solutions, which prevents the model to be equivalent to
standard quantum cosmology. Furthermore, the topological solution dominates at the classical
limit. We also study the effect of an Immirzi parameter in the model.

\end{abstract}

\newpage

\section*{1. Introduction}
Among the issues which remain to be understood in Loop Quantm Gravity (LQG) \cite{LQG}, the problem of the 
dynamics is surely one of the most important. The regularization of the Hamiltonian constraint proposed by
Thiemann \cite{Th} was a first promissing attempt towards a solution of that problem. However, the technical
difficulties are such that this way has not given a solution yet. Spin Foam models \cite{Pe} is an alternative
way to explore the question: they are supposed to give a combinatorial expression of the Path integral of
gravity and then they should allow to compute transition amplitudes between states of quantum gravity, or equivalently
to compute the dynamics of a state.
For a long time, the Barrett-Crane model \cite{BC} 
has been the only one Spin-Foam model a priori
relevant for quantum gravity: it was introduced as a quantization of Plebanski theory \cite{Pl}
which is a $BF$ theory \cite{BF} where the $B$ field is constrained to be simple.
Classically, the Plebanski theory contains the solutions of first order gravity.
The Barrett-Crane model has been deeply studied but its relation to the LQG dynamics 
remains mysterious. Recently, a new model has been proposed \cite{Ca} whose link with LQG is much more
transparent: it is expected to give the vertex of LQG or in other words the matrix elements of the
quantum Hamiltonian constraint. 
This model is very promising and is under intensive study to see whether it satisfies all the
properties it should. 

Face with such difficulties (to solve the problem of dynamics), simplified models of LQG have been
considered, the most popular being the Loop Quantum Cosmology (LQC) introduced by Bojowald \cite{Bo}.
This model appears much richer than one could expect: it gives a solution to the problem of the cosmological
singularity (that does not exist anymore in this model) and it proposes a framework to test important issues 
related to the dynamics of the full LQG. 

When we see the success of LQC, it seems quite natural to
consider the cosmological Plebanski model, i.e. the cosmological symmetry reduction of the
Plebanski theory, to test some issues related to loop quantum gravity and spin-foam models. 
This is exactly what we do in this paper. More precisely, we perform the classical analysis 
of the cosmological Plebanski model to clarify its link with standard cosmology. Then, we quantize 
the theory both canonically and covariantly to compare to two schemes. In the canonical framework, we recover
the standard space of states which consists on the sole Kodama state \cite{Ko}. The path integral quantization,
on the contrary, is quite different from what would be a path integral of standard cosmology. The reasons of these
discrepencies are the following: at the classical level, Plebanski theory contains solutions of first order
gravity but also the so-called topological solutions which are physically irrelevant; these solutions are classically
distinct but the path integral mixed them and the resulting amplitude is different from what would be
an amplitude for quantum cosmology. Furthermore, the topological solutions dominate at the classical limit.
One way to overcome the problem is to modify the Plebanski theory such that it does not admit the topological
solutions anymore. There is a way to do so \cite{CaMo} but the price to pay is that 
classical solutions are those of gravity with an Immirzi parameter $\gamma$ supplemented with those
of gravity with the inverse parameter $\gamma^{-1}$. Thus, the path integral can be performed and can be shown,
as expected, to mix both types of solutions. This prevents the quantum theory to be equivalent to a quantum theory
of cosmology. These results leads naturally to the question of the legitimity to consider the Plebanski theory
as a starting point to understand quantum gravity and spin-foam models. 

\medskip

The paper is organized as follows. Next section is devoted to the classical theory. We start by recalling
the basic properties of the full Plebanski theory in the presence of an Immirzi parameter. Then, we perform
the cosmological symmetry reduction and make the complete Hamiltonian analysis. Section 3 is devoted to the
quantization of the model both canonically and covariantly. In the canonical framework, we found the physical
solutions and make the contact with the standard loop quantum cosmology results. In the covariant framework,
we compute the path integral and shows that the obtained amplitude is quite different from the standard amplitude
for quantum cosmology. When there is no Immirzi parameter, we show that the topological sector dominates at
the classical limit. We finish with some discussions.

\section*{2. The Classical Model}
In this section, we will construct the cosmological Plebanski model imposing homogeneity and isotropy to the full Plebanski theory \cite{Pl}. First, we will briefly review basic notions about Plebanski theory. Afterwards, we will present the symmetry reduction and analyse (classically) the obtained model.

In what follows, the space-time is a four dimensional smooth oriented manifold ${\cal M}$. We will assume that its topology is given by ${\cal M} = \Sigma \times \mathbb R$ where the space $\Sigma$ is a three-dimensional manifold and the real line $\mathbb R$ stands for time direction. For our purposes, $\Sigma$ could be an ``open space'' ($\mathbb R^3$) or a ``closed space'' (the sphere $S^3$).

\subsection*{2.1. Plebanski theory}
The Plebanski theory is a constrained $BF$ theory \cite{BF}
in the sense that the (so called) $B$-field is enforced to be simple by the simplicity constraints. Given a Lie Group $G=SO(4)$ or $G=SO(3,1)$ (respectively for the Riemannian or Lorentzian theory), whose Lie algebra is denoted $\mathfrak g$, its action is given by a functional $S[A,\Sigma,\varphi]$ of the two-form $\Sigma$ with  value in the Lie algebra, the connection $A$ and the Lagrange multipliers $\varphi$. In the following, we will use greek letters ($\mu,\nu \cdots$) to denote space-time indices, capital latin letters ($I,J \cdots \in \{0,1,2,3\}$) for ($\mathfrak g$-) internal  indices, small latin letters from the begining of the alphabet $(a,b \cdots)$ for space indices, $t$ denotes the time label (in the $\mathbb R$ direction) and small latin letters from the middle of the alphabet $(i,j \cdots \in \{1,2,3\})$ for ($sl(2)$-) internal indices.

Let us note that there exists two formulations of this action depending whether $\varphi$ is a density tensor field with space-time indices only or a field with algebra (internal) indices only. The two formulations are equivalent (in the non-degenerate sector) but, for our convenience, we will adopt the latter which is furthermore the one used to study spin-foam models \cite{Pe}. 
Thus, we will consider the following expression for the Plebanski action (with a cosmological term $L$ which will be related to the usual cosmological constant latter):
\begin{eqnarray}\label{Full Plebanski action}
S[A,\Sigma,\varphi]= \frac{1}{2} \int_{\cal M} \! Tr(\Sigma \wedge F(A) + \frac{L}{2} \star \Sigma \wedge \Sigma+\varphi(\Sigma) \wedge \Sigma)
\end{eqnarray}
As usual $F(A)=F_{\mu \nu}(A) dx^{\mu} \wedge dx^{\nu}$ is the curvature of the connection $A$; $Tr$ is a Killing form in the Lie algebra; $\star$ is the Hodge map of the Lie algebra (note that $\star^2=\sigma^2$ where the parameter $\sigma$ takes the value $1$ or $i$ depending on the gauge group $SO(4)$ or $SO(3,1)$) and we have introduced the notation $\varphi(\Sigma)^{IJ}=\varphi^{IJ}{}_{KL} \Sigma^{KL}$. The Lagrange multiplier $\varphi_{IJKL}$ is symmetric into the exchange of the pairs $[IJ]$ and $[JK]$, satisfies the traceless condition $\epsilon^{IJKL} \varphi_{IJKL} = 0$ and therefore admits $20$ independent components and enforces the field $\Sigma$ to be simple. Non degenerate solutions of these constraints have been classified into a gravitationnal and a topological sectors \cite{Fr}. 
Restricted to the gravitationnal sector, the theory is equivalent to first-order gravity. The canonical analysis of the Plebanski theory have been performed (in the other version) which is a first step toward an eventual Hamiltonian quantization \cite{He}.

If one replaces the traceless condition on the Lagrange multiplier by the more general condition:
\begin{eqnarray}\label{General traceless condition}
\mu^{IJKL} \varphi_{IJKL} \; = \; 0 \;\;\;\text{with}\;\;\;\; \mu_{IJKL} = \epsilon_{IJKL} + 2\zeta \; \eta_{IK} \eta_{JL}\;\;,
\end{eqnarray}
$\zeta$ being a non-zero real parameter, then the theory (restricted to a suitable sector of solutions) is equivalent to gravity supplemented with an Immirzi parameter $\gamma$, solution of the algebraic equation $2\zeta^{-1} = \gamma^{-1} + \sigma^2 \gamma$.  Note that if $\gamma$ is a solution of the previous equation so is $\sigma^2 \gamma^{-1}$; therefore the theory (\ref{Full Plebanski action}) describes two sectors of general relativity which differ by the value of the Immirzi parameter and also by the value of the cosmological constant $\Lambda$ which depends on $\gamma$ as follows $\Lambda=-L(1 + \sigma^2/\gamma^2)$. Classically, the sectors are disjoint but the quantization will mix them as we will see in the cosmological case.

\medskip

In order to study simultaneously the Riemannian and Lorentzian cases, it will be convenient to complexify the theory keeping at hands reality conditions on the dynamical fields. The complexified Plebanski action takes the same form as the real action (\ref{Full Plebanski action}) but the gauge group is the complex group $G_{\mathbb C}=SO(4,\mathbb C)$ and the dynamical variables become complex. The Lagrange multipliers $\varphi$ and the parameter $\zeta$ in the relation (\ref{General traceless condition}) are also a priori complex variables.

The complex Lie algebra $\mathfrak g_{\mathbb C}=so(4,\mathbb C)$ is a direct sum of two commuting $sl(2,\mathbb C)$ algebras. Therefore any element $\xi$ of the complex Lie algebra $\mathfrak g_{\mathbb C}$ can be written as a sum of its self-dual component $\xi^{{}^{{}_{(+)}}}$ and its anti-self-dual one  $\xi^{{}^{{}_{(-)}}}$. The original real Lie algebra is recovered when one imposes reality conditions on the generators of the complex Lie algebra (to select the compact or non-compact real form). Applying this decomposition to the connection $A$, to the field $\Sigma$ and to the Lagrange multiplier $\varphi \in \mathfrak g \otimes \mathfrak g$ (the $20$ independent components of the Lagrange multiplier $\varphi$ are decomposed into $2$ symmetric traceless $3\times 3$ matrices $\phi_{ij}^{{}^{_{(\pm)}}}$, a scalar $\phi$ and a $3 \times 3$ matrix $\psi_{ij}$), one can show that the complex Plebanski action takes the following form:
\begin{eqnarray}\label{self dual action}
S[A,\Sigma,\varphi] & = & S^{(+)}[A,\Sigma,\varphi] \; + \; S^{(-)}[A,\Sigma,\varphi] \; + \; I[\Sigma,\varphi]
\end{eqnarray}
where we have defined the actions:
\begin{eqnarray}
S^{(\pm)}[A,\Sigma,\varphi] & = & \frac{1}{2} \int_{\cal M}\!\! Tr(\Sigma^{{}^{{}_{(\pm)}}} \!\wedge F^{{}^{{}_{(\pm)}}}_{{}^{{}}} \!+\frac{\phi^{{}^{{}_{(\pm)}}} \!\!\pm \sigma L}{2} \; \Sigma^{{}^{{}_{(\pm)}}} \!\wedge \Sigma^{{}^{{}_{(\pm)}}}_{{}^{{}}}) + \frac{1}{2} \phi_{ij}^{(\pm)} \; \Sigma^{{}^{{}_{(\pm)i}}} \!\wedge \Sigma^{{}^{{}_{(\pm)j}}} \nonumber \\
I[\Sigma,\varphi] & = & \frac{1}{2} \int_{\cal M}\! \psi_{ij} \; \Sigma^{{}^{{}_{(+)i}}} \wedge \Sigma^{{}^{{}_{(-)j}}} \;\; .\nonumber
\end{eqnarray}
We have introduced the notation $\phi^{{}^{{}_{(\pm)}}}\!\!=\phi(1\mp \frac{\zeta}{\sigma})$ and $F^{{}^{_{(\pm)}}}$ stands for the curvature of the self-dual or anti-self-dual part of the connection. The real theory is recovered when one imposes reality conditions on the dynamical fields induced by reality conditions on the complex Lie algebra:
\begin{eqnarray}\label{Reality conditions}
\overline{A_{\mu}^{{}^{{}_{(\pm)i}}}} \; = \; A_{\mu}^{{}^{{}_{(\sigma^2\pm)i}}}\;\;\; \text{and} \;\;\; \overline{\Sigma_{\mu \nu}^{{}^{{}_{(\pm)i}}}} \; = \; \Sigma_{\mu \nu}^{{}^{{}_{(\sigma^2\pm)i}}}\;.
\end{eqnarray}
When $\zeta$ is a real parameter, one sees immediately from previous reality conditions that the action (\ref{self dual action}) is real.

\subsection*{2.2. Symmetry reduction and Lagrangian analysis}
Symmetry reduction of gravity has been considered and extensively studied in several formulations of general relativity. In particular, cosmological reduction in terms of Ashtekar variables has been carried out and developped by 
Bojowald \cite{Bo}. 
In this section we adapt methods he has developped to define the cosmological reduction of Plebanski theory.

\subsubsection*{2.2.1. Symmetry reduction}
The cosmological Plebanski model is defined by imposing spacial homogeneity and isotropy to the full theory (\ref{Full Plebanski action}). Using ``Loop quantum cosmology'' notations, we consider the one-forms $\omega^I = \omega_a^I dx^a$ (and their dual vector fields $\omega^a_I$) which are left-invariant with respect to the translational symmetry associated with homogeneity and we introduce the left-invariant space metric $q^0_{ab} = \omega^I_a \omega^J_b \delta_{IJ}$. Therefore, components of any homogeneous and isotropic spatial $G-$connections read:
\begin{eqnarray}\label{Cosmological connection}
A^{{}^{{}_{(\pm)i}}}_a \!\! = \; A^{{}^{{}_{(\pm)}}} \Lambda^{{}^{{}_{(\pm)}}}{}^i_I \; \omega^I_a \;\;\; \text{with} \;\;\;\;\;\;\; \overline{\Lambda^{{}^{{}_{(\pm)}}}{}^i_I}=\Lambda^{{}^{{}_{(\sigma^2\pm)}}}{}^i_I\;.
\end{eqnarray}
As a result, the gauge invariant part of the connection is encoded in the parameters $A^{{}^{{}_{(\pm)}}}$. The $SO(3,\mathbb C)$ matrices $\Lambda^{{}^{{}_{(\pm)}}}{}^i_I$ caracterise the gauge dependent part of the connection; the conditions on their components (\ref{Cosmological connection}) are induced by reality conditions which select the real form of the gauge group. Note that $A_t$ vanishes due to isotropy.

In order to reduce the field $\Sigma$ to homogeneity and isotropy, we first separate its components into $B_a^{{}^{{}_{(\pm)i}}} = \Sigma_{0a}^{{}^{{}_{(\pm)i}}}$ and $\tilde{E}{}^a{}{{}^{{}_{(\pm)i}}} = \epsilon^{abc} \Sigma_{bc}^{{}^{{}_{(\pm)i}}}$. Then, applying the same techniques as above (for the connection), one shows immediately that the gauge invariant parts of the field $\Sigma$ are encoded in parameters $B^{{}^{{}_{(\pm)}}}$ and $\tilde{E}{}^{{}^{{}_{(\pm)}}}$ as follows:
\begin{eqnarray}\label{Cosmological two-form}
B^{{}^{{}_{(\pm)i}}}_a \!\! = \; B^{{}^{{}_{(\pm)}}} \Lambda^{{}^{{}_{(\pm)}}}{}^i_I \; \omega^I_a \;\;\;\;\;\; \text{and} \;\;\;\;\;\; \tilde{E}{}{}^a{}^{{}^{{}_{(\pm)i}}}\!\! = \; \tilde{E}{}^{{}^{{}_{(\pm)}}} \Lambda^{{}^{{}_{(\pm)}}}{}_j^I \delta^{ij} \; \omega^a_I \;.
\end{eqnarray}

If we replace the connection $A$ and the field $\Sigma$ by their reduced expressions given by (\ref{Cosmological connection}, \ref{Cosmological two-form}) in the full Plebanski action, we obtain (up to a global volume factor $ 3 V^0 = 3\int_{\Sigma} \!d^3x \sqrt{\det[q^0]}$ that we can eliminate by absorbing it in a redefinition of the dynamical variables) the action for the cosmological Plebanski model:
\begin{eqnarray}\label{Cosmological Plebanski action}
S[A,B,E;\phi,\psi] \; = \; \int \! dt \sum_{\varepsilon = \pm}(E^{{}^{{}_{(\varepsilon)}}} \!\dot{A}{}^{{}^{{}_{(\varepsilon)}}} \! + 2B^{{}^{{}_{(\varepsilon)}}}\![A^{{}^{{}_{(\varepsilon)}}}{}^2 - n A^{{}^{{}_{(\varepsilon)}}}]) + L {\cal V} + \phi \chi_1 \!+ \psi \chi_2
\end{eqnarray}
where we have introduced the notations $E^{{}^{{}_{(\pm)}}} = \det[q^0]^{-1/2} \tilde{E}{}^{{}^{{}_{(\pm)}}}$, $\psi=3\psi_{ii}$ and:
\begin{eqnarray}\label{Cosmological simplicity constraints}
{\cal V} & = & \sigma(E^{{}^{{}_{(+)}}}B^{{}^{{}_{(+)}}}-E^{{}^{{}_{(-)}}}B^{{}^{{}_{(-)}}}) \;\;,\\
\chi_1 & = & (1-\sigma \zeta)E^{{}^{{}_{(+)}}}B^{{}^{{}_{(+)}}}+(1+\sigma \zeta)E^{{}^{{}_{(-)}}}B^{{}^{{}_{(-)}}} \;\;,\\
\chi_2 & = & E^{{}^{{}_{(+)}}}B^{{}^{{}_{(-)}}}+E^{{}^{{}_{(-)}}}B^{{}^{{}_{(+)}}}\;\;.
\end{eqnarray}
The parameter $n=0$ for an isotropic flat space ($\Sigma = \mathbb R^3$) and $n=1$ for an isotropic closed space ($\Sigma=S^3$) and $\cal V$ is the expression of the space-time volume expressed in Plebanski variables. At this point, the theory is a priori complex and we recover the real theory by imposing the following reality conditions on the ``cosmological'' degrees of freedom (which are a straightforward consequence of (\ref{Reality conditions})):
\begin{eqnarray}\label{Cosmological reality conditions}
\overline{A^{{}^{_{(\pm)}}}} \; = \; A^{{}^{_{(\sigma^2\pm)}}} \;\;\; , \;\;\; \overline{B^{{}^{_{(\pm)}}}} \; = \; B^{{}^{_{(\sigma^2\pm)}}} \;\;\; \text{and} \;\;\; \overline{E^{{}^{_{(\pm)}}}} \; = \; E^{{}^{_{(\sigma^2\pm)}}} \;\;.
\end{eqnarray}
Therefore, the cosmological Plebanski model is completely defined by the complex action (\ref{Cosmological Plebanski action}) supplemented with the previous reality conditions. As for the full Plebanski theory, it is a constrained theory; in particular, the set of simplicity constraints reduces to a set of only two constraints given by $\chi_1 \simeq 0$ and $\chi_2 \simeq 0$ (\ref{Cosmological simplicity constraints}).

\subsubsection*{2.2.2. Classical solutions and link to cosmology}
To make contact with the usual formulation of cosmology, let us first solve the simplicity constraints. Solutions are classified into degenerate and non-degenerate ones. A solution is said degenerate when the space-time volume vanishes on this solution, i.e. ${\cal V}=0$; otherwise it is non-degenerate. One can show that a degenerate solution which satisfies reality conditions (\ref{Cosmological reality conditions}) is such that $E^{{}^{_{(\pm)}}} \!\!= 0$ or $B^{{}^{_{(\pm)}}} \!\!= 0$. As a consequence, any non degenerate solution is such that $E^{{}^{_{(\pm)}}}$ and $B^{{}^{_{(\pm)}}}$ are non-zero functions of time. Note that if we naively extrapolate to the case $\zeta = \pm \sigma$, the constraint $\chi_1$ admits only degenerate solutions. In fact, the case $\zeta = \pm \sigma$ is technically different because the action becomes non-singular and does not produce second class constraints anymore; it corresponds to the self-dual Plebanski model and has been studied in detail in \cite{Ke}. In  the sequel, we will concentrate only on non-degenerate solutions and therefore we will assume that $\zeta \neq \pm \sigma$. Such solutions satisfy the following conditions:
\begin{eqnarray}\label{Non degenerate conditions}
E^{{}^{_{(+)}}} \!\! = - q \; E^{{}^{_{(-)}}} \; \text{and} \;\; B^{{}^{_{(+)}}} \!\! =  q \; B^{{}^{_{(-)}}}\;\text{where} \;\; q\;\;\text{is a solution of} \;\; q^2 = \frac{1+\sigma \zeta}{1-\sigma \zeta}\;.
\end{eqnarray}
Due to the reality conditions, the parameter $q$ is a pure phase in the Lorentzian regime and a real number in the Riemannian regime. Moreover, one can see that, for the Riemannian theory to be consistent, the parameter $\zeta$ must belong to the open ball $]-1,+1[$ (this is a constistency condition between simplicity constraints and reality conditions). There is no such a restriction in the Lorentzian model. In the case where $q \neq -1$ (therefore $\zeta \neq 0$), the conditions (\ref{Non degenerate conditions}) are equivalent to the existence of non-null functions $E$ and $B$ such that:
\begin{eqnarray}\label{Non degenerate solutions}
E^{{}^{_{(\pm)}}} \!\! = \mp\sigma E(1 \pm \frac{\sigma}{\gamma}) \;\;\;\; \text{and} \;\;\;\; B^{{}^{_{(\pm)}}} \!\! = B(1 \pm \frac{\sigma}{\gamma}) \;\;\; \text{where} \;\;\; \gamma = \sigma \frac{q+1}{q-1} \;.
\end{eqnarray}
Note that the parameter $\gamma$ is real in both Lorentzian and
Riemannian models. Besides, from the reality conditions
(\ref{Cosmological reality conditions}), one can show that
$\overline{E}= E$ and $\overline{B}= B$. Therefore $E$ and $B$ are
real whatever the regime (Lorentzian or Euclidean) we consider.

If one injects non-degenerate solutions (\ref{Non degenerate solutions}) into the original action (\ref{Cosmological Plebanski action}), one obtains that the dynamics of $E$, $B$ and the components $A^{{}^{_{(\pm)}}}$ of the connection is governed by an action of the form:
\begin{eqnarray}\label{Cosmological Holst model}
S[A,B,E]  =  S_0[A,B,E] \; + \; \frac{\sigma}{\gamma} \; S_1[A,B,E]
\end{eqnarray}
where $S_0[A,B,E]$ and $S_1[A,B,E]$ correspond respectively to the standard first order action for cosmology and a (cosmological) topological term whose Lagrangian are respectively given by:
\begin{eqnarray}
{\cal L}_0 & = &  -\sigma E(\dot{A}{}^{{}^{_{(+)}}}\!\! - \dot{A}{}^{{}^{_{(-)}}}) + 2B(-\sigma^2 L(1+\frac{\sigma^2}{\gamma^2}) E + A^{{}^{_{(+)}}}{}^2 + A^{{}^{_{(-)}}}{}^2 -nA^{{}^{_{(+)}}} - n A^{{}^{_{(-)}}}) \nonumber\\
{\cal L}_1 & = &  -\sigma E(\dot{A}{}^{{}^{_{(+)}}} \!\!+ \dot{A}{}^{{}^{_{(-)}}}) + 2B(A^{{}^{_{(+)}}}{}^2 - A^{{}^{_{(-)}}}{}^2 -nA^{{}^{_{(+)}}} + n A^{{}^{_{(-)}}}) \nonumber
\end{eqnarray}
Note that we have recovered the true value of the cosmological constant in the expression of $S_0$. In fact, the action (\ref{Cosmological Holst model}) is the cosmological analogue of the Holst action \cite{Ho} 
for first order gravity where $\gamma$ is the Immirzi parameter. Indeed, if we first solve the equations of motion for the components of the connection, we obtain that
$A^{{}^{_{(\pm)}}} \!\! = \! \frac{1}{2} (n \mp \sigma \frac{\dot{E}}{2B})$;
then, if we replace these expressions in the action (\ref{Cosmological Holst model}), we see immediately that the topological part of the action vanishes identically whereas the term $S_0$ gives back the usual second order cosmological action, i.e.:
\begin{eqnarray}
S_{cos}[E,B] \; = \; \int \!\! dt \;\left( -\frac{\sigma^2}{4B}\dot{E}{}^2 + B(2\sigma^2 \Lambda E - n^2 ) \right)\;\;.
\end{eqnarray}
The value of the cosmological constant is then fixed by $\Lambda= -L(1+\frac{\sigma^2}{\gamma^2}) $. The topological term does not modify the dynamics of the classical model as in the full theory and, when $\zeta \neq 0$, we show that the model (\ref{Cosmological Plebanski action}) is classically equivalent to standard cosmology. However, the conclusions of our analysis are deeply modified when the Immirzi parameter vanishes, which corresponds to $\zeta = 0$. In that case, there exist two inequivalent sectors of non-degenerate solutions of the simplicity constraints:
\begin{enumerate}
\item the gravitational sector: $E^{{}^{_{(+)}}} = -E^{{}^{_{(-)}}} = E$ and $B^{{}^{_{(+)}}} = B^{{}^{_{(-)}}} = B$;
\item the topological sector: $E^{{}^{_{(+)}}} = E^{{}^{_{(-)}}} = E$ and $B^{{}^{_{(+)}}} = - B^{{}^{_{(-)}}} = B$.
\end{enumerate}
These sectors are the cosmological analogues of those present in the full Plebanski theory. In particular, restricted to the gravitational sector, the model is equivalent to standard cosmology (as above) whereas it has no physical meanning when restricted to the topological sector. As a consequence, we see that the presence of an Immirzi parameter in the Plebanski model ``eliminates'' the topological sector. We will precise this property when we perform the canonical analysis of the model in the next subsection.

\subsection*{2.3. Hamiltonian analysis}
We start with the action (\ref{Cosmological Plebanski action}) and we assume for the moment that $\zeta \in \mathbb C -\{0,\pm \sigma\}$. The (non-physical) phase space  is 8-dimensional and parametrised by the canonical variables defined by the following non-vanishing Poisson brackets:
\begin{eqnarray}\label{Non physical phase space}
\{E^{{}^{_{(\pm)}}},A^{{}^{_{(\pm)}}}\}\; = \; 1 \;\;\;\;\; \text{and} \;\;\;\;\; \{B^{{}^{_{(\pm)}}},P^{{}^{_{(\pm)}}}\} \; = \; 1 \;.
\end{eqnarray}
In the sequel, we will restrict the phase space to the non-degenerate sector, i.e. we will assume that the variables $E^{{}^{_{(\pm)}}}$ and $B^{{}^{_{(\pm)}}}$ are not identically null. 

\subsubsection*{2.3.1. Constraints analysis}
The variables $\phi$ and $\chi$ are Lagrange multipliers and therefore are not considered as dynamical variables of the theory. The phase space (\ref{Non physical phase space}) is singular and the theory admits the primary constraints $\chi_1 \simeq 0$, $\chi_2 \simeq 0$ and $P^{{}^{_{(\pm)}}} \simeq 0$. Therefore, the total Hamiltonian $H_T$ reads:
\begin{eqnarray}
- H_T \; = \; {\cal{H}} \; + \; \phi \chi_1 \; + \; \psi \chi_2  \; + \; \lambda^{{}^{_{(+)}}} P^{{}^{_{(+)}}} \; + \; \lambda^{{}^{_{(-)}}} P^{{}^{_{(-)}}}
\end{eqnarray}
where ${\cal H} = 2B^{{}^{_{(+)}}}(A^{{}^{_{(+)}}}{}^2 - n A^{{}^{_{(+)}}}) + 2B^{{}^{_{(-)}}}(A^{{}^{_{(-)}}}{}^2 - n A^{{}^{_{(-)}}}) + L {\cal V}$ and we have introduced the Lagrange multipliers $\lambda^{{}^{_{(\pm)}}}$ to enforce the primary constraints $P^{{}^{_{(\pm)}}}\simeq 0$. Time derivative of any phase space function $f$ is given by $\dot{f} = \{f,H_T\}$. Conservation of simplicity constraints implies the relations:
\begin{eqnarray}
\dot{\chi}_1 \!\!& = &\!\!\!\! -2B^{{}^{_{(+)}}}{}^2(2A^{{}^{_{(+)}}} - n) -2q^2 B^{{}^{_{(-)}}}{}^2(2A^{{}^{_{(-)}}} - n) - \lambda^{{}^{_{(+)}}} E^{{}^{_{(+)}}}\!\! -\!\! q^2 \lambda^{{}^{_{(-)}}} E^{{}^{_{(-)}}} \simeq 0\\
\dot{\chi}_2 \!\!& = &\!\!\!\! -4B^{{}^{_{(+)}}}B^{{}^{_{(-)}}}(A^{{}^{_{(+)}}} + A^{{}^{_{(-)}}} - n) - \lambda^{{}^{_{(-)}}} E^{{}^{_{(+)}}} - \lambda^{{}^{_{(+)}}} E^{{}^{_{(-)}}} \simeq 0
\end{eqnarray}
These two relations combine together to, first, fix one Lagrange multiplier in term of the other and also to give a new secondary constraint as follows:
\begin{eqnarray}\label{Constraint T}
B^{{}^{_{(-)}}} \dot{\chi}_1 \; + \; B^{{}^{_{(+)}}} \dot{\chi}_2 \; \simeq \; 0 \;\;\; \Longrightarrow \;\;\; T \equiv A^{{}^{_{(+)}}} + A^{{}^{_{(-)}}} - n \; \simeq \; 0 \;.
\end{eqnarray}
In the same way, time conservation of the primary constraints $P{}^{{}^{_{(\pm)}}} \simeq 0$ gives two equations involving the Lagrange multipliers $\lambda^{{}^{_{(\pm)}}}$:
\begin{eqnarray}
\dot{P}{}^{{}^{_{(\pm)}}}  & = &  2(A^{{}^{_{(\pm)}}}{}^2-nA^{{}^{_{(\pm)}}}) \pm \sigma L E^{{}^{_{(\pm)}}} + q^{1\mp1}\phi E^{{}^{_{(\pm)}}} + \psi E^{{}^{_{(\mp)}}} \simeq 0 \;.
\end{eqnarray}
One can show, as previously, that these equations fix only one of the two Lagrange multipliers and one gets ${\cal H} \simeq 0$ as a new secondary constraint using simply the fact that ${\cal H} \simeq B^{{}^{_{(+)}}}\dot{P}{}^{{}^{_{(+)}}} +  B^{{}^{_{(-)}}}\dot{P}{}^{{}^{_{(-)}}}$. One has to continue the Dirac algorithm by imposing conservation of the secondary constraints under time evolution. It appears that $\dot{\cal H}=0$ only fixes Lagrange multipliers and does not impose new tertiary constraints. Time derivative of the constraint (\ref{Constraint T}) gives:
\begin{eqnarray}\label{Last condition}
\dot{T} \; = \; -\phi (B^{{}^{_{(+)}}} - q^2 B^{{}^{_{(-)}}}) - \psi(B^{{}^{_{(+)}}} + B^{{}^{_{(-)}}}) - \sigma L (B^{{}^{_{(+)}}} - B^{{}^{_{(-)}}})\; \simeq \; 0 \;.
\end{eqnarray}
In the case of a non-vanishing Immirzi parameter (i.e. $q \neq 0$), then time conservation of $T$ does not imply new constraints and therefore the Dirac algorithm closes. However, the conclusion is subtler when the Immirzi parameter vanishes because it depends on the sector of solutions we are considering. Indeed, in the gravitational sector, equation (\ref{Last condition}) fixes the Lagrange multipliers, we do not have more constraints and the Dirac algorithm closes. In the topological sector, (\ref{Last condition}) does not fix Lagrange multipliers anymore and imposes $B^{{}^{_{(+)}}} - B^{{}^{_{(-)}}} \simeq 0$ as a new tertiary constraint (if the cosmological constant is non-null). As a result, the components $B^{{}^{_{(+)}}}$ and $B^{{}^{_{(-)}}}$ weakly vanish, therefore the topological sector is degenerate and the theory becomes trivial in that case.

\subsubsection*{2.3.2. Dirac bracket}
We continue with the case of a non-vanishing Immirzi parameter. The system admits six constraints that we have to split into first class and second class. To achieve this aim, we first check that $\chi_1, \chi_2, \chi_3 \equiv T$ and $\chi_4 \equiv P^{{}^{_{(+)}}} - P^{{}^{_{(-)}}}$ form a set of second class constraints: we compute their Dirac matrix $\Delta$ (i.e. the matrix of the Poisson brackets of the constraints $\Delta_{ij} =\{\chi_i,\chi_j\}$) and show its invertibility. A direct calculation shows:
\begin{eqnarray}
\Delta = \left( \begin{array}{cc} 0 &  M \\ -M^t & 0 \end{array} \right)\;\;\; \text{where} \;\;\; M = \left( \begin{array}{cc} B^{{}^{_{(+)}}} \!\!+ q^2 B^{{}^{_{(-)}}} &  E^{{}^{_{(+)}}} \!\!- q^2 E^{{}^{_{(-)}}} \\ B^{{}^{_{(+)}}}+ B^{{}^{_{(-)}}}  & E^{{}^{_{(-)}}} \!\!- E^{{}^{_{(+)}}} \end{array} \right)
\end{eqnarray}
The determinant is easy to compute and is on shell proportional to the space-time volume $\cal V$:
\begin{eqnarray}
\det{\Delta}\; = \; \det{M}^2 \;\;\; \text{with} \;\;\; \det{M} \; = -2 \frac{(1+q)^2}{q} B^{{}^{_{(+)}}}E^{{}^{_{(+)}}}\; (\propto \; {\cal V}) \;.
\end{eqnarray}
Therefore, the Dirac matrix $\Delta$ is manifestly invertible (in the non-degenerate sector) and the constraints $(\chi_i)_{i=1,\cdots,4}$ form a set of second class constraints. Furthermore, one can show (see below) that the remaining constraints ${\cal H}$ and ${\cal S}=P^{{}^{_{(+)}}} + P^{{}^{_{(-)}}}$ are first class (up to second class constraints); then, as one could expect, the physical phase space is zero dimensional and consists on a single state. The Dirac bracket between any two functions $f$ and $g$ on the phase space reads:
\begin{eqnarray}\label{Dirac bracket}
\{f,g\}_D \; = \; \{f,g\} - \{f,\chi_i\}\Delta^{ij}\{\chi_j,g\}
\end{eqnarray}
where we have denoted by $\Delta^{ij}$ the coefficients of the inverse of the Dirac matrix. In particular, one shows that:
\begin{eqnarray}
\{E^{{}^{_{(+)}}},A^{{}^{_{(+)}}}\}_{D} \; = \; \frac{\gamma+\sigma}{2\gamma} \; = \; \{B^{{}^{_{(+)}}},P^{{}^{_{(+)}}}\}_D \;.
\end{eqnarray}
As we have the explicit expression of the Dirac bracket, one can solve the second class constraints. The constraints surface is four-dimensional and is locally parametrized by the functions $E$, $B$, $\Omega$ and $P$ related to the original variables by:
\begin{eqnarray}\label{Expression of Omega}
E^{{}^{_{(\pm)}}}\!\!=-\sigma E(\sigma \gamma \pm 1) , \; B^{{}^{_{(\pm)}}}\!\!=B(1\pm \sigma \gamma)  , \; 2 A^{{}^{_{(\pm)}}} \!\!= (1 \mp \sigma \gamma)n \mp \frac{1}{\sigma}\Omega , \; P^{{}^{_{(\pm)}}}\!\! = \frac{1}{2} P.
\end{eqnarray}
The constraints surface inherits a symplectic structure from the Dirac bracket (\ref{Dirac bracket}) and we show that $(E,\Omega)$ and $(B,P)$ are pairs of canonical variables, i.e.:
\begin{eqnarray}\label{Physical phase space}
\{E,\Omega\}_D \; = \; 1 \;\;\;\; \text{and} \;\;\;\; \{B,P\}_D \; = \;1\;.
\end{eqnarray}
These are the only non-vanishing Dirac brackets on the basic variables. At this level, the phase space is a priori complex and we have to impose reality conditions to obtain the real section: all canonical variables are real.
Finally, physical classical states are points of the real constraints surface which satisfy first class constraints ${\cal H}\simeq 0$ and ${\cal S}\simeq 0$ up to symmetries (generated by the first class constraints themselves). In fact, the theory admits only one physical state as expected. The constraint ${\cal S}\simeq 0$ imposes the fact the $B$ is not a dynamical variable and therefore can be sent to zero. The constraint $\cal H$ is the only one remaining and can be written in terms of the variables (\ref{Physical phase space}) as follows:
\begin{eqnarray}\label{Cosmological hamiltonian}
\nonumber {\cal H}[\Omega,E] & = & \sigma^2 B ((\Omega
+\frac{\sigma^2 n}{\gamma} )^2 - \sigma^2 n^2
-2L(1+\frac{\sigma^2}{\gamma^2}) E)\\ 
& = &  \sigma^2 B ((\Omega
+\frac{\sigma^2 n}{\gamma} )^2 - \sigma^2 n^2 +2\Lambda E) \;.
\end{eqnarray}

\subsubsection*{2.3.3. Link to canonical gravity}
Note that this constraint is the symmetry reduced analogue of the
Hamiltonian constraint of gravity. Indeed, if we first start from
the full theory, then recall that self-dual and anti-self-dual
components of the connections are expressed in term of the
intrinsic curvature $K_\mu^i$ and the Christoffel symbol
$\Gamma_\mu^i$ and finally apply the cosmological symmetry
reduction, we show that our variables are related to usual gravity
variables by:
\begin{eqnarray}
A_\mu^{{}^{_{(\pm)}}}{}^j \! = \; \Gamma_{\mu}^j \pm \sigma K_\mu^j  \;\;\; \Longrightarrow \;\;\; A^{{}^{_{(\pm)}}} \! = \; \frac{1}{2}(n \pm \sigma K) \;,
\end{eqnarray}
where $\Gamma = \frac{n}{2}$ and $K$ are the gauge-invariant parts of the tensors $\Gamma_\mu^i$ and $K_\mu^i$. Therefore, the ``connection'' $\Omega$ is related to the intrinsic curvature by the relation $\Omega=-\frac{\sigma^2 n}{\gamma} -2\sigma^2 K$ and there exists a trivial canonical transformation from the phase space variables $(E,\Omega)$ to the variables $(2E,-\sigma^2 K)$ in term of which the Hamiltonian constraint (\ref{Cosmological hamiltonian}) reads:
\begin{eqnarray}\label{ADM constraint}
{\cal H}[K,\sigma^2 E] \; = \; 4\sigma^2 B (K^2 - \sigma^2 \Gamma^2 + \frac{L}{2}(1+\frac{\sigma^2}{\gamma}) E \; .
\end{eqnarray}
Up to some eventual global factors, this expression is manifestly the (cosmological) symmetry reduced version of the ADM Hamiltonian. This result holds also in the case of a vanishing Immirzi parameter as one could expect.

To recover complex Ashtekar self dual formulation of gravity, one has to send the parameter $\gamma$ to $\sigma$ in the action (\ref{Cosmological Holst model}). Therefore, the anti-self dual component of the connection disapears from the action; there is no more second class constraints but we have to deal with reality conditions which take the form $A^{{}^{_{(\pm)}}} + \overline{A^{{}^{_{(\pm)}}}} = \frac{1-\sigma^2}{2} n$. This condition is a simplified version of the well-known reality conditions in the full theory and is completely solvable. Thus, we end up with a real physical phase space which is shown to be trivially related to the previous one ($\gamma \in \mathbb R$) by a canonical transformation. Moreover, the Hamiltonian constraint in the self-dual formulation takes exactly the same form as (\ref{ADM constraint}) once reality conditions are solved.

Finally, the gauge invariant part of the Ashtekar-Barbero connection ${}^{\gamma} A$ is given by
\begin{eqnarray}\label{Barbero variables}
 {}^{\gamma} A \; = \; \Gamma + \gamma K \; = \; -\frac{\gamma \sigma^2}{2} \Omega.
\end{eqnarray}
There is a trivial canonical transformation from $(E,\Omega)$ to $(-\frac{2\sigma^2}{\gamma}E,{}^\gamma A)$ generated by the function $F=\theta E(K+ \gamma n)$ where we have introduced the infinitesimal parameter $\theta=-\log \gamma$ (for positive $\gamma$). The Hamiltonian constraint (\ref{ADM constraint}) can be trivially expressed in terms of Ashtekar-Barbero variables.

Before going to the quantization, let us recall basic results concerning the clasical analysis of our model. The (non degenerate) classical phase space consists in a disjoint union of the symplectic spaces, both corresponding to gravity phase space but differing by their values of the Immirzi parameter: $\gamma$ and $\frac{\sigma^2}{\gamma}$. Therefore, the presence of the Immirzi parameter eliminates the topological sector in the cosmological Plebanski model. This property is still valid in the quantum theory but the quantization is going to ``mix'' the two phase spaces, allowing for transitions between the two spaces. We will precise the very meanning of that in the sequel.

\section*{3. The Quantum Theory}
This section aims at quantizing the cosmological Plebanski model with a real Immirzi parameter. First, we consider a canonical quantization and make contact with loop quantum cosmology. Then, we study the path integral quantization which should mimic the ``spin foam'' quantization of Plebanski theory: in that framework, we explore the issue of the measure and we compute the semi-classical limit. Finally, we compare the two quantization schemes.

\subsection*{3.1. ``Loop'' or Canonical Quantization}
We want first to quantize the reduced phase space (\ref{Physical
phase space}) and then to implement at the quantum level the
Hamiltonian constraint (\ref{Cosmological hamiltonian}).

We choose the polarization where the connection $\Omega$ is the
configuration variable. Therefore, states in this representation
are wave functions of the connection $\Psi(\Omega)$. We promote
the connection $\Omega$ and the frame field $E$ respectively to
multiplicative and derivative operators acting on the wave
functions. We endow the vector space of states with the usual
Hilbert structure on real functions and normalizable states are
elements of $L^2(\mathbb R)$.

\subsubsection*{3.1.1. The physical state of Kodama}
Physical states are solutions of the Hamiltonian constraint
(\ref{Cosmological hamiltonian}). In fact, there exists only one
state (for each sector) which is a pure phase (up to a global
constant) given by:
\begin{eqnarray}\label{Physical solution}
\Psi_{phy}^{\gamma}(\Omega)\; = \; \exp\lbrace \frac{i}{2\hbar
\Lambda(\gamma)}[\frac{1}{3}(\Omega + \sigma^2 \gamma n)^3 -
\sigma^2 n^2 \Omega]\rbrace\;,
\end{eqnarray}
where we explicitly write the dependence of
$\Lambda(\gamma)=-L(1+\sigma^2/\gamma^2)$ as this will be
important for what follows. Note that this state does not define a
$L^2(\mathbb R)$ function and therefore it is not normalizable
with the kinematical scalar product. However, the state is
delta-normalizable in the sense defined by Freidel and Smolin,
whatever the signature (Euclidean or Lorentzian) of our model \cite{FS}.

Moreover, this state should be related to a quantization of deSitter space-time. In fact, it is trivially related to the Kodama state \cite{Ko}
when one reduces the self-dual connection to homogeneity and isotropy. The Kodama state is a special solution of self-dual quantum gravity in the presence of a positive cosmological constant and is given by the expression:
\begin{eqnarray}\label{Kodama}
\Psi_K(A^{{}^{_{(+)}}}) \; = \; \exp \frac{3i\sigma}{\lambda} \int_{\Sigma} d^3x \;( \epsilon^{abc} A^{{}^{_{(+)}}}_a{}^i \partial_b A^{{}^{_{(+)}}}_{ci} + \frac{1}{3}\epsilon_{ijk} A^{{}^{_{(+)}}}_a{}^i A^{{}^{_{(+)}}}_b{}^j A^{{}^{_{(+)}}}_c{}^k)\;.
\end{eqnarray}
We have introduced the dimensionless constant $\lambda=G\hbar \Lambda$. It is interesting to note that this wave function is effectively a solution of euclidean quantum gravity but we have to impose the reality conditions to obtain a complete solution in the lorentzian regime. No one knows how to implement reality conditions in general. But these conditions are much simpler to implement in the cosmological model and one obtains the cosmological ``Kodama'' solution  in the lorentzian regime (\ref{Physical solution}). To make a clear contact between the two expressions (\ref{Physical solution}) and (\ref{Kodama}), we start by replacing in (\ref{Kodama}) the self-dual connection by the homogeneous and isotropic self-dual connection (\ref{Cosmological connection}) and we obtain after a few calculation (and a rescaling by the space volume $V^0$) that:
\begin{eqnarray}\label{Cosmological Kodama solution}
\Psi_K(A^{{}^{_{(+)}}}) \; = \; \exp \frac{3i\sigma}{\lambda}(2A^{{}^{_{(+)}}}{}^3 -3n A^{{}^{_{(+)}}}{}^2)
\end{eqnarray}
Finally, we replace the gauge invariant part of the self-dual connection by its expression in terms of $\Omega$ (\ref{Expression of Omega}) and we obtain a function which is proportional to the state (\ref{Physical solution}). Therefore, the physical state we found is strickly the Kodama state.

It is interesting to underline, once again, that we obtain a
complete solution in the Euclidean and Lorentzian regimes. Let us
concentrate on the Lorentzian solution. One can see the expression
(\ref{Cosmological Kodama solution}) as a solution of cosmological
self-dual quantum gravity before implementing the reality
conditions: it is a real exponential and clearly divergent with
respect to the ``kinematical'' scalar product (i.e. the usual Haar
measure on the complex plane). It is even not delta integrable as
the euclidean solution is. Reality conditions are very simple and
reads $A^{{}^{_{(+)}}} + \overline{A^{{}^{_{(+)}}}}=2n$.
Therefore, imposing the reality conditions is equivalent to change
the scalar product by replacing the integral on the whole complex
plane with the integral on the line $A^{{}^{_{(+)}}} +
\overline{A^{{}^{_{(+)}}}}=2n$. Then, we can deform the contour of
integration to be over the real line. Doing this, we find the
lorentzian physical solution which is still not integrable but at
least it behaves exactly in the same way as the euclidean one.
This simple model shows that one can solve successfully reality
conditions and obtain physical well-behaved solutions.

\subsubsection*{3.1.2. Link to loop quantum cosmology}
We have proposed above a ``direct'' canonical quantization of the cosmological Plebanski model in the sense that we have implemented the usual Dirac procedure to quantize the Hamiltonian constraint. Loop quantum cosmology starts from a slightly different perspective: its aim is to mimic as close as possible the techniques developed in the full theory to test them in the framework of cosmology \cite{Bo}.

The first step required is to build the kinematical Hilbert space: it is defined by square integrable functions on $\mathbb R_{Bohr}$, the so-called Bohr compactification of the real line. The set of almost periodic functions defines an orthonogonal basis of this Hilbert space. We come naturally to this Hilbert space if, using ideas from the full theory, we define the classical elementary variables to be matrix elements of holonomies of the connection along edges.

Then, we have to solve the Hamiltonian constraint (\ref{Cosmological hamiltonian}) to construct the physical Hilbert space from the kinematical one. The problem is that the Hamiltonian constraint  is a ill-defined operator in the kinematical Hilbert space because no operator corresponding to the connection $\Omega$ exists. In fact, connection operators need to be represented in terms of the basic variables; i.e. matrix element of holonomies along edges. In particular, the curvature of the connection can be approximated by the holonomy of the connection along a closed loop. Because of isotropy, we can choose holonomies around squares with edges length chosen to be $\nu_0 V_0^{1/3}$ for some positive parameter $\nu_0$. Following these ideas, one obtains immediately the expression of the regularized Hamiltonian constraint ${\cal H}_{\nu_0}$:
\begin{eqnarray}
{\cal H}_{\nu_0}[\Omega,E] \; = \; (\frac{\sin(\nu_0 \Omega)}{\nu_0})^2 + 2\Lambda E \; \simeq \; 0\;.
\end{eqnarray}
For purposes of simplicity, we have considered only the case of flat cosmology ($n=0$). It is clear that the structure of the Hamiltonian constraint has drastically changed after the regularisation and the solution $\Psi^{\nu_0}_{phy}$ is no longer given by the Kodama state but is straightforward to derive:
\begin{eqnarray}
\Psi^{\nu_0}_{phy}(\Omega) \; = \; \exp\{\frac{i}{4\Lambda \hbar \nu_0^2}[\Omega + \frac{\sin(2\nu_0 \Omega)}{2\nu_0}] \}\;.
\end{eqnarray}
Note that we could have work with Ashtekar-Barbero variables (\ref{Barbero variables}) instead; but there is an obvious ambiguity introduced by the regularization procedure which simply does not commute with the canonical transformation (mapping Ashtekar variables to Ashtekar-Barbero ones). This results from the fact that one is using holonomies instead of connection variables and therefore the canonical transformation is no longer an unitary transformation at the level of the quantum theory.

In standard loop quantum cosmology, solutions of the Hamiltonian constraint in the connection represention does not exist because the Hamiltonian constraint is more involved. Hence, one works in the triad representation. In our model, we can also construct the physical Hilbert space starting from the triad representation. At the kinematical level, states are labelled by a real parameter $\nu$ and the Hamiltonian constraint leads to a difference equation that physical solutions have to satisfy. This point has been studied in details in \cite{Ke}. In particular, we have shown that the classical singularity does not present any obstruction for the evolution of the universe, as it did
in standard loop quantum cosmology.

\subsection*{3.2. ``Spin-Foam'' or Path Integral Quantization}
The aim of this section is to perform the path integral quantization of the cosmological Plebanski model. For that purpose, we need to define the path integral measure before performing the integration. The measure has been  computed in the full theory \cite{He}. In the previous section, we have also computed the measure for the reduced model from the classical canonical analysis. Therefore, the path integral is given by the formula:
\begin{eqnarray}\label{General path integral}
{\cal Z} \; = \; \int [{\cal D}A^{{}^{_{(\pm)}}}] [{\cal D}E^{{}^{_{(\pm)}}}] [{\cal D}B^{{}^{_{(\pm)}}}] [{\cal D}P^{{}^{_{(\pm)}}}][{\cal D}\mu] \prod_t \sqrt{\det(\Delta)}\prod_{\alpha=1}^4 \delta(\chi_{\alpha}) \exp \frac{i}{\hbar} S\;.
\end{eqnarray}
We will restrict the integral on the non-degenerate sector and we will consider for the moment the Euclidean model. Recall the expression of the action $S$:
\begin{eqnarray}
S\; = \; \int \! dt \sum_{\varepsilon = \pm}(E^{{}^{{}_{(\varepsilon)}}} \!\dot{A}{}^{{}^{{}_{(\varepsilon)}}} \! + 2B^{{}^{{}_{(\varepsilon)}}}\![A^{{}^{{}_{(\varepsilon)}}}{}^2 - n A^{{}^{{}_{(\varepsilon)}}}]) + L {\cal V} + \mu P^{{}^{_{(+)}}} \;.
\end{eqnarray}
Note that the variable $\mu$ is a Lagrange multiplier which enforces the first class constraint $P^{{}^{_{(+)}}}\simeq 0$.

\subsubsection*{3.2.1. Simplification of the path integral expression}
The first step consists on simplifying this expression in order to recover the path integral defined on the reduced phase space. During this procedure, we will show the importance of the presence of the measure in the definition of the path integral. We start by writting the constraints $\chi_1$ and $\chi_2$ as follows:
\begin{eqnarray}
\chi_1 = \frac{1}{2}(C_E^+ C_B^+ + C_E^-C_B^-) \;\; , \;\; \chi_2 = \frac{1}{2q}(C_E^+ C_B^+ - C_E^-C_B^-) \\
\text{with}\;\;\;\; C_E^{\pm} = E^{{}^{{}_{(+)}}} \pm qE^{{}^{{}_{(-)}}} \;\;\; \text{and}\;\;\; C_B^{\pm} = B^{{}^{{}_{(+)}}} \pm qB^{{}^{{}_{(-)}}}\;.
\end{eqnarray}
In the non-degenerate sector, the functions $C_E^+$ and $C_E^-$ cannot vanish simultaneitly and the same property holds for $C_B^+$ and $C_B^-$. Therefore, we can write the equality
\begin{eqnarray}
\delta(\chi_1) \cdot \delta(\chi_2) \; = \; 2\vert q \vert \left( \frac{\delta(C_E^+) \cdot \delta(C_B^-)}{\vert C_E^- \vert \cdot \vert C_B^+\vert} + \frac{\delta(C_E^-) \cdot \delta(C_B^+)}{\vert C_E^+ \vert \cdot \vert C_B^-\vert}\right)
\end{eqnarray}
we can use to simplify the expression of the path integral (\ref{General path integral}) which becomes (after some integrations):
\begin{eqnarray}
&&{\cal Z} =   \int [{\cal D}A^{{}^{_{(-)}}}] [{\cal D}E^{{}^{_{(-)}}}] [{\cal D}B^{{}^{_{(-)}}}] [{\cal D}P^{{}^{_{(-)}}}][{\cal D}\mu] \\ \nonumber
&&\hspace{2cm} \int[{\cal D}E^{{}^{_{(+)}}}][{\cal D}B^{{}^{_{(+)}}}] \prod_t (\frac{\delta(C_E^+)\cdot \delta(C_B^-)}{(1+q)^{-2}} + \frac{\delta(C_E^-)\cdot \delta(C_B^+)}{(1-q)^{-2}}) \exp \frac{i}{\hbar} S \;.
\end{eqnarray}
We can formally integrate over the variables $E^{{}^{_{(+)}}}$ and $B^{{}^{_{(+)}}}$ and we see that two weights contribute to the path integral. We emphasize the fact that the measure of the path integral has been cancelled exactly after integration. Therefore, not considering the measure in the definition of the path integral would have introduced a singular term in the integral such that the degenerate sector would have dominated.

The integrations over $P^{{}^{_{(-)}}}$ and $\mu$ are immediate and a simple variables changing (\ref{Expression of Omega}) gives:
\begin{eqnarray}\label{Cosmological path integral}
&&{\cal Z} = \int [{\cal D}\Omega] [{\cal D}E][{\cal D}B] \int \prod_t[{\cal D}x(t)] (\delta(x-\gamma) + \delta(x+\frac{\sigma^2}{\gamma}))\exp \frac{i}{\hbar}S_x[A,E,B]  \;,
\end{eqnarray}
where we have introduced the action:
\begin{eqnarray}\label{Action for path integral}
S_x[A,E,B] \; = \; \int dt \left( E \dot\Omega -\sigma^2 B((\Omega+nx)^2 - \sigma^2 n^2 - 2L(1+\sigma^2x^2)E) \right)\;.
\end{eqnarray}
As expected,we recover the action of cosmology. Note that the path
integral has two contributions which correspond to the two
possible values of the Immirzi parameter $\gamma$ and $\sigma^2
\gamma^{-1}$.  Indeed, the starting point was the Plebanski action
with modified simplicity constraints (\ref{General traceless
condition}); classicaly this theory was shown to be equivalent (in
the non-degenerate sector) to gravity supplemented with an Immirzi
parameter $\gamma$ whose value is fixed by the equation
$\frac{2}{\zeta}=\gamma + \frac{\sigma^2}{\gamma}$. Therefore, if
$\gamma$ is a solution, $\sigma^2\gamma^{-1}$ is also a solution.
From this moment on we will refer to the two sectors as $q$ and
$-q$ respectively (see equation \ref{Non degenerate solutions}).
At the classical level, these two sectors are obviously disjoint.
However, at the quantum level, there are mixed and there could
exist interferences between the two sectors.

For the Lorentzian model, we have to be more careful even if we
adopt technically the same strategy to simplify the expression of
the path integral. The main difference is that the integration
variables in (\ref{General path integral}) are complex variables.
Therefore we have to impose the reality conditions (in the path
integral) expliciting the fact that $A^{{}^{_{(+)}}}$ and
$A^{{}^{_{(-)}}}$ are complex conjugate and so on. The second
class constraints impose to restrict the integrations (initially
on the whole complex plane) to the real line (after some trivial
contour deformation). Finally, we end up with the same result as
the one given above (\ref{Cosmological path integral}) with the
Lorentzian Hamiltonian in the action (\ref{Action for path
integral}). This expression of the path integral is the starting
point to define a spin foam model for cosmological Plebanski
model.

\subsubsection*{3.2.2. Physical scalar product with Immirzi parameter}
Spin-foam models were introduced as an eventual way to compute the
physical scalar product for loop quantum gravity. So far, no clear
connection between spin-foam models and canonical loop quantum
gravity has been done. 
Recently new models have been considered where the link between
the two quantization becomes much more transparent \cite{Ca}.
These new models are very promissing.
In principle, a spin-foam model should give
a discretization of the path integral of gravity and therefore
should represent a way to compute amplitude transitions between
physical states. Unfortunately, spin-foam models are just at the
level of a program and were never used to compute an explicit
amplitude transition. The simple model we are proposing is an
exemple where we can illustrate some of these ideas.

Kinematical states are described as functions of the connection
$\Omega$ (in the connection representation) and form an Hilbert
space whose Hilbert structure is given by the usual $\mathbb
R$-invariant measure. We will denote by $d\mu(\Omega)$ this
measure. The physical scalar product can be computed from the path
integral. If we denote by $P(\Omega',\Omega)$ the matrix elements
of the path integral (the propagator in the connection
representation), then the physical scalar product
$<\phi,\psi>_{Phy}$ between two kinematical states $\phi$ and
$\psi$ is given formally by the formula:
\begin{eqnarray}
<\phi,\psi>_{Phy} \; = \; \int d\mu(\Omega)d\mu(\Omega') \; \overline{\phi(\Omega)} P(\Omega',\Omega) \psi(\Omega)\;.
\end{eqnarray}
From the expression of the path integral (\ref{Cosmological path
integral}), we can write the propagator $P(\Omega',\Omega)$ as the
sum:
\begin{eqnarray}
P(\Omega',\Omega) \; = \; K_{q}(\Omega',\Omega) \; + \;
K_{-q}(\Omega',\Omega) \; + \; I(\Omega',\Omega)\;.
\end{eqnarray}
The kernel $K_x$ corresponds to the propagator of gravity in the
presence of an Immirzi parameter $x$ and $I$ represents the
interference kernel between the two gravitational sectors
(corresponding to values of the parameters $q$ and $-q$). The
propagator of self-dual Plebanski theory (reduced to cosmology)
has already been computed in \cite{Ke}. From there it follows that
$K_q$ \footnote{If we ignore the existence of the other sector
then the projection kernel would simply be given by
\[K_q\;= \;\ket{\  q\ }\bra{\ q\ } \;,\]
and similarly for the other sector. However, in the path integral
we must sum over all paths, including those for which the system
jumps to the $-q$ sector and comes back to the the $q$ sector.
This sum can be organized in such a way that the amplitude of the
full sum becomes
\[K_q\;= \;\ket{\ q\ }\bra{\ q\ } +\ket{\ q\ }\bra{\ q \ }\!\!\ket{-q}
\bra{\ q \ }\!\!\ket{-q}\bra{\ q\ }+\ket{\ q\ }\bra{\ q \
}\!\!\ket{-q} \bra{\ q \ }\!\!\ket{-q}\bra{\ q \ }\!\!\ket{-q}
\bra{\ q \ }\!\!\ket{-q}\bra{\ q\ }+\cdots,\] where the first term
is the sum over all paths where no transition to the $-q$ sector
is made, the second terms corresponds to all the paths where the
systems jumps to $-q$ and comes back and so on. If we define
$a=\bra{\ q\ }\!\!\!\; \ket{-q}$ we simply get
\[K_q\;= \;\ket{\ q\ }\bra{\ q\ } (1+a\overline{a}+(a\overline{a})^2+\cdots),\]
which corresponds to expression in the main text if we re-sum the
series in the standard way.}:
\begin{eqnarray}\label{Pure gravity propagator}
K_q\;= \; \frac{1}{1-a\overline{a}}\ \ket{\  q\ }\bra{\ q\ } \;,
\end{eqnarray}
while the interference term yields
\begin{eqnarray}\label{Pure gravity propagator}
I \;= \; \frac{ a \ \ket{\ q\ }\bra{-q} +
\overline{a}\ \ket{\! -q }\bra{\ q\ }}{a\overline{a}-1} \;,
\end{eqnarray}
where we are using Dirac notation to write the solutions of the
Hamiltonian constraint (\ref{Physical solution}) for the sectors
$q$ (or simply $\gamma$) and $-q$ (or $\sigma^2\gamma^{-1}$)
respectively. The complex number $a$ is defined as
\[a=\bra{\ q\ }\!\!\!\; \ket{-q}=\int d\mu(\Omega) \
\overline{\Psi_{phy}^{q}}(\Omega)\ \Psi_{phy}^{-q}(\Omega). \] It
is easy to see that $a$ is well defined. From the expression of
the physical inner product we conclude that even though classical
we are dealing with two sectors $q$ and $-q$, quantum mechanically
there is non vanishing quantum interference between the sectors
$q$ and $-q$. In essence what has happened is that 
the value of $q$ has become dynamical in the quantum theory, and
hence that of the cosmological constant. In fact if we use the
obvious bracket Dirac notation the projector becomes:
\begin{eqnarray}\label{propo}
P=\frac{\ket{\ q\ } \bra{\ q\ }+\ket{-q } \bra{-q}- a \ \ket{\ q\
} \bra{-q }-\overline{a} \ \ket{- q } \bra{\ q\
}}{1-a\overline{a}}
\end{eqnarray}
It is easy to check that the previous expression simply
corresponds to the identity operator by orthonormalizing the basis
$\ket{q}$ and $\ket{-q}$ using the Gram-Scchmith method. So the
physical Hilbert space is genuinely $2$-dimensional. The syatem
has degrees of freedom at the quantum level!

Once we realize that $P=1$ then the probability interpretation is
straightforward. In fact $p_{\gamma\rightarrow
\sigma^2\gamma^{-1}}=a\overline{a}$ is the transition probability
form one sector to the other.
\begin{equation}
\overline \Lambda=-
\frac{L}{2}(1+\sigma^2\frac{1+\gamma^4}{\gamma^2}).
\end{equation}
All this can be interpreted in a very simple way if we look at the
form of the projector (\ref{propo}). In fact the physical Hilbert
space has in fact dimension one, and is spanned by the ray defined
by the state $\ket{\gamma}+\ket{{\vani \sigma}^{\va 2}\gamma^{\va
-1}}$. The other orthogonal combination of solutions of the
Hamiltonian constraint, namely $\ket{\gamma}-\ket{\sigma^{\va
2}\gamma^{\va -1}}$ has in fact zero physical norm.


\subsubsection*{3.2.3. Physical scalar product without Immirzi parameter}
We could have studied cosmological Plebanski theory without
Immirzi parameter but the situation is in fact more complicated in
that case: the Hamiltonian system is non-regular in the sense that
the rank of the Dirac matrix is non-constant on the non-reduced
phase space. If we perform the canonical analysis, we have to
distinguish between the gravitational and the topological sector.
The phase space restricted to the gravitational sector the same as
the phase space (\ref{Physical phase space}) where the Hamiltonian
is given by (\ref{Cosmological hamiltonian}) when the Immirzi
parameter $\gamma$ is sent to infinity. The topological phase
space of cosmological Plebanski theory is also the same but the
Hamiltonian is given by $H=EB$. The two sectors contribute to the
path integral and therefore the propagator (in the connection
representation) can be written as:
\begin{eqnarray}
P(\Omega',\Omega) \; = \; K_{\infty}(\Omega',\Omega) \; + \;
K_{0}(\Omega',\Omega) \; + \; K_{\infty \rightarrow
0}(\Omega',\Omega) \;.
\end{eqnarray}
The term $K_{\infty}$ is the gravitational propagator (\ref{Pure
gravity propagator}) with null Immirzi parameter, the topological
propagator $K_{0}$ is given by
$K_0(\Omega',\Omega)=\delta(\Omega'-\Omega)$ and $K_{\infty
\rightarrow 0}$ denotes the interference term between the
gravitational (with $\gamma=\infty$) and topological sectors
($\gamma=0$). In that sense, Plebanski theory is not, at the
quantum level, equivalent to gravity. It is well known that, even
if two theories admit the same classical solutions (in some
sectors), there could admit inequivalent quantization. This is
exactly what happens in the case presented here. But, Plebanski
theory could be a good starting point to quantize gravity if, at
least, the gravitational sector dominates at the semi-classical
limit.

In the simple model at hand, we can compute the relative weight
between the gravitational and topological sectors to see which
dominate at the semi-classical limit. To do so, let us compute the
transition amplitude between two kinematical states $\phi$ and
$\psi$. The ratio $Q(\phi,\psi)$ between the gravitational and the
topological contributions is given by:
\begin{eqnarray}\label{ratio}
Q(\phi,\psi) \; = \; \frac{<\phi,\psi>_G}{<\phi,\psi>_T}\;,
\end{eqnarray}
where $<\phi,\psi>_G$ and $<\phi,\psi>_T$ denote respectively the gravitational and the topological contributions given by:
\begin{eqnarray}
<\phi,\psi>_G & = & \int d\mu(\Omega) d\mu(\Omega') \; \overline{\phi(\Omega)} P_0(\Omega,\Omega') \psi(\Omega')  \\
<\phi,\psi>_T & = & \int d\mu(\Omega) \; \overline{\phi(\Omega)} \; \psi(\Omega) \;.
\end{eqnarray}
There is no Planck constant in the topological contribution and therefore its classical limit is itself. The Planck constant appears in the gravitational contribution and a straightforward analysis shows that:
\begin{eqnarray}
<\phi,\psi>_G \; = \; (6\hbar \Lambda)^{\frac{2}{3}} \; I^2 \; \overline{\phi(0)} \cdot \psi(0) \; + \; O(\hbar) \;\; \text{where} \;\; I = \int_{-\infty}^{+\infty} dx \; e^{ix^3} \;.
\end{eqnarray}
Generically (when $<\phi,\psi>_T \neq 0$), the topological sector clearly dominates at the classical limit and the ratio (\ref{ratio}) is given by:
\begin{eqnarray}
Q = (6\hbar \Lambda)^{\frac{2}{3}} \; I^2 \; \frac{\overline{\phi(0)} \cdot \psi(0)}{\int  d\mu(\Omega) \; \overline{\phi(\Omega)} \; \psi(\Omega)}\; + O(\hbar) \;.
\end{eqnarray}
Therefore, the quantization of the cosmological Plebanski theory (with no Immirzi parameter) does not give back general relativity at the classical limit.

\section*{4. Conclusion}
This paper is devoted to an extensive study of the cosmological Plebanski model, namely the
cosmological symmetry reduced version of Plebanski theory. In the absence of
Immirzi parameter, classical solutions are classified into two different and disjoint sectors
(as for the full theory):
the topological one which is not physically relevant and the gravitational one which consists 
in the classical solutions of cosmology. In the presence of an Immirzi parameter, there is no more
topological sector but two gravitational sectors, both corresponding to solutions of classical cosmology, one 
with an Immirzi parameter $q$ and the other with a parameter $-q$. This is also very similar to what happens in 
the full Plebanski theory. Thus, as expected
the cosmological Plebanski  model is quite different from standard cosmology for it contains more 
classical solutions. However, physically irrelevant solutions can be easily identified and, in that sense, eliminated.

This has been known since a long time in the full theory.
The reason to consider cosmology symmetry reduction is that we can perform the quantization to understand 
the role of the topological sector. In the canonical quantization, one can easily distinguish gravitational from
topological solutions: in the gravitational sector, the Hilbert space consists in a sole state which is the 
cosmological symmetry reduced Kodama state. The path integral quantization mixes, as expected, both sectors
which prevents the obtained amplitude to be the one of quantum cosmology. The situation is even worse because
the topological sector dominates at the classical limit. One way to eliminate the physically irrelevant solutions
is introducing an Immirzi parameter in the model: the path integral of such a model is not the one of quantum
cosmology. The reason, recalled above, is that it mixes two theories of gravity with different Immirzi parameter.

\medskip

Of course, one cannot extrapolate these results to the full theory. Nevertheless, one can ask the
question of the legitimity to use Plebanski theory as a starting point for quantum gravity (in the context
of spin-foam models). Is it possible to find a way to eliminate the topological sector while quantizing?
This question is of course very complicated but certainly deserves to be studied.

\bibliographystyle{unsrt}

\end{document}